\documentclass[a4paper,10pt,twoside]{cpc-hepnp}

\usepackage{multicol}
\usepackage{graphicx}
\usepackage{booktabs}
\usepackage{amssymb,bm,mathrsfs,bbm,amscd}
\usepackage[tbtags]{amsmath}
\usepackage{lastpage}

\begin{document}

\fancyhead[c]{\small Chinese Physics C~~~Vol. 39, No. 5 (2015)056003} \fancyfoot[C]{\small 056003-\thepage}

\footnotetext[0]{Received 26 August 2014, Revised 8 November 2014}

\title{A Cosmic Ray Test Platform Based on the High Time Resolution MRPC Technology\thanks{Supported by the National Natural Science Foundation of China under Grant No. 11275196, and the ¡°Strategic Priority Research Program¡± of the Chinese Academy of Sciences under Grant No. XDA03010100}}

\author{%
      CHEN Tian-Xiang³ÂÌìÏè$^{1;2)}$\email{licheng@ustc.edu.cn}%
\quad LI ChengÀî³Î$^{1;1)}$\email{ctx@mail.ustc.edu.cn}%
\quad SUN Yong-JieËïÓ½Ü$^{1}$\\
\quad CHEN Hong-Fang³Âºê·¼$^{1}$
\quad SHAO MingÉÛÃ÷$^{1}$
\quad TANG Ze-BoÌÆÔó²¨$^{1}$\\
\quad YANG Rong-XingÑîÈÙÐÇ$^{1}$
\quad ZHOU YiÖÜÒâ$^{1}$
\quad ZHANG Yi-FeiÕÅÒ»·É$^{1}$
}
\maketitle

\address{%
$^1$ State Key Laboratory of Particle Detection and Electronics, University of Science and Technology of China, Hefei 230026, China \\
}

\begin{abstract}
In order to test the performance of detector/prototype in environment of laboratory, we design and build a larger area ($90\times52$ cm$^2$) test platform of cosmic ray based on well-designed Multi-gap Resistive Plate Chamber (MRPC) with an excellent time resolution and a high detection efficiency for the minimum ionizing particles (MIPs). The time resolution of the MRPC module used is tested to be $\backsim$80 ps, and the position resolution along the strip is $\backsim$5 mm, while the position resolution perpendicular to the strip is $\backsim$12.7 mm. The platform constructed by four MRPC modules can be functional for tracking the cosmic rays with a spatial resolution $\backsim$6.3 mm, and provide a reference time $\backsim$40 ps.
\end{abstract}

\begin{keyword}
Test platform,  MRPC,  Time resolution,  Cosmic ray tracking
\end{keyword}

\begin{pacs}
29.20.Ej \qquad  \textbf{DOI:}10.1088/1674-1137/39/5/056003
\end{pacs}

\footnotetext[0]{\hspace*{-3mm}\raisebox{0.3ex}{$\scriptstyle\copyright$}2013
Chinese Physical Society and the Institute of High Energy Physics
of the Chinese Academy of Sciences and the Institute
of Modern Physics of the Chinese Academy of Sciences and IOP Publishing Ltd}%

\begin{multicols}{2}

\section{Introduction}

Cosmic ray is a natural and inexhaustible source for the test of the particle detectors in the laboratory. Cosmic ray test platform is essential and useful for understanding the detector performance, in particular which can provide the response of the detector to the MIPs. A test platform with large sensitive area and high spatial resolution can precisely track the cosmic rays with large acceptance. As a gaseous detector working in avalanche mode, MRPC has an excellent time resolution and a high detecting efficiency [1]. In addition, MRPC can be constructed into large area modules with long readout strips without decreasing its performance. The incident position can be achieved with the time difference of signals from each end of the strip [2, 3]. Therefore, MRPC is a very good choice for building a large area test platform of cosmic ray. The MRPC module used in this paper is one of the well-designed prototypes developed by USTC group for the RHIC-STAR Muon Telescope Detector (MTD) [4]. For tracking the cosmic ray, four similar MRPC modules are stacked to measure the impact positions of the cosmic ray passed through them, where the time and position resolution of each detector can be measured simultaneously. In this paper, we describe the details of the test platform based on the high time resolution MRPC technology and present some results obtained with the muon events from cosmic ray.

\section{The MRPC module}

The left panel of Fig. 1 schematically shows the structure of the MRPC module with 5 gas gaps. The gap is defined to be 250 $\mu$m by nylon fishing line. The resistive plates are made of 0.7 mm float glass sheets with volume resistivity of $\backsim 10^{13}$ $\Omega \cdot $cm. The inner glass is 1.6 cm shorter than outermost glass. The electrodes are made of graphite coating sprayed on the outer glasses.

\begin{center}
\includegraphics[width=8cm]{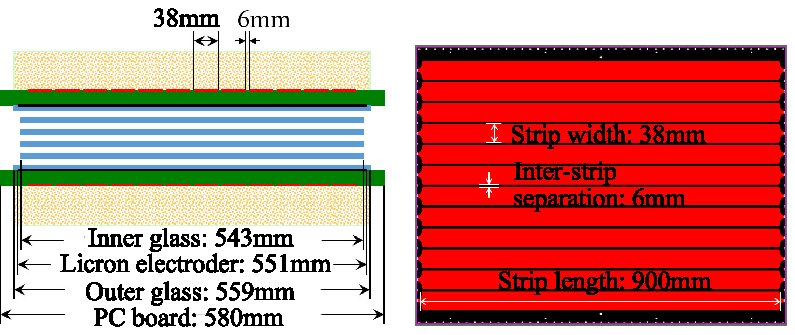}
\figcaption{\label{fig1}Left panel: Schematic side-view of the MRPC module. Right panel: The readout printed circuit board.}
\end{center}

The read out electrodes are segmented into twelve double-ended strips, which are 38 mm wide and 90 cm long, as shown in the right panel of Fig. 1. The gap between each strip is 6 mm wide. The active area of this MRPC module is $90.0 \times 52$ cm$^2$. For the readout, differential signals are obtained from the anode and cathode pickup strips at both ends, and sent to the front-end electronics (FEE). The hit position along the strip can be calculated from the time difference. The MRPC modules are enclosed in a gas tight aluminum box. The working gas mixture contains $95\%$ $Freon$ $R-134a,$ $4.5\%$ $iso-C_4H_{10}$ and $0.5\%$ $SF_6$.

\begin{center}
\includegraphics[width=8cm]{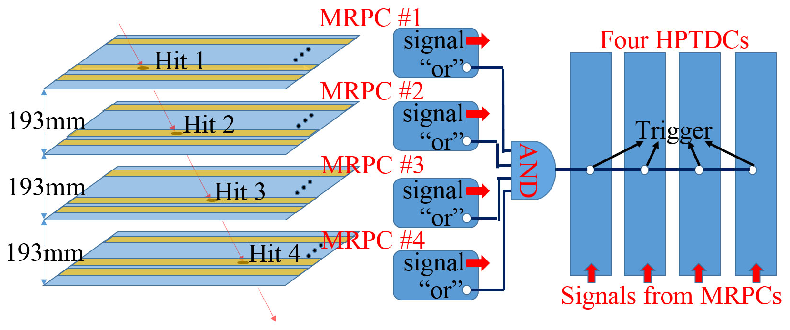}
\figcaption{\label{fig1}Left panel: Schematic view of cosmic ray set up. Right panel: Trigger and DAQ.}
\end{center}

\section{Experiment setup}
Four MRPC modules are stacked as shown in Fig. 2 left panel. A 6-channel amplifier/shaping FEE card based on the NINO chip is used to connect the readout strips [5, 6]. The leading edge of the output pulse, a Low-Voltage Differential Signaling (LVDS) signal, offers the precise timing of the signal. The width of LVDS represents the amplitude of the signal. The FEE card also produces an OR output of all 6 channels. If an OR signal received from both ends of one module, we assume a muon has passed the module. The logical scheme is shown in Fig. 3.

\begin{center}
\includegraphics[width=8cm]{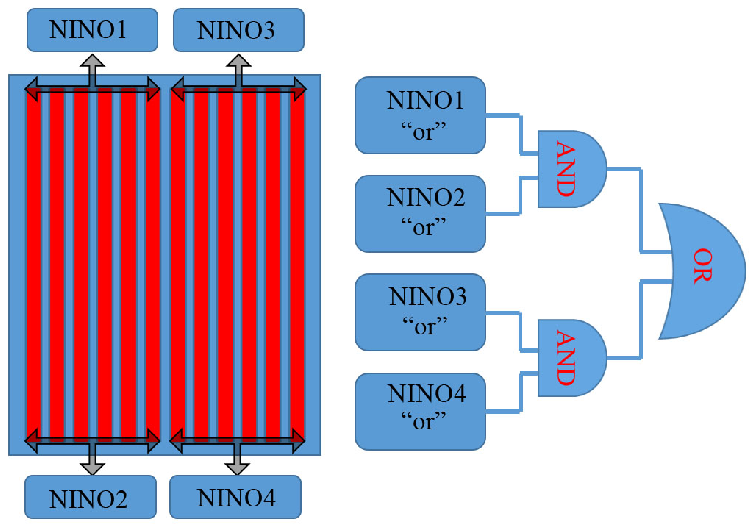}
\figcaption{\label{fig1}The logical scheme of the ¡°or¡± from each MRPC.}
\end{center}

The data acquisition (DAQ) system should be triggered when the muon go through all of the four modules, as shown in Fig. 2 right panel. The DAQ system is a LabVIEW program running on a PC, which is connected to a VME crate via a CAEN V1718 USB-VME bridge. The TDCs (CAEN V1290A)used for readout data from MRPCs is a 1-unit wide VME 6U module that houses 32 independent Multi-Hit /Multi-Event Time to Digital Conversion channels with resolution of 25 ps.

\section{Testing and analyzing}

\subsection{The detecting efficiency and time resolution}

The MRPC's plateau of detecting efficiency and the time resolution versus high voltage (HV) are shown in Fig. 4. At higher HV, the statistic of avalanche signals is more stable, which makes better time resolution. But, after $\pm$6.4 kV, the increasing streamer signals will make the time resolution getting worse. We chose the working HV at $\pm$6.3 kV. The corresponding detecting efficiency is about $90\%$, while the time resolution of the MRPC module is $\backsim$80 ps after TOT and path length correction, the detail of which are presented in the following paragraphs.

\begin{center}
\includegraphics[width=6cm]{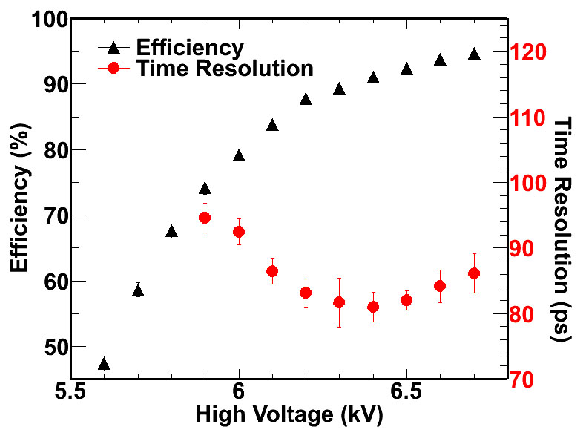}
\figcaption{\label{fig2}Detecting efficiency and time resolution versus HV.}
\end{center}

\subsubsection{The detecting efficiency measurement}

In the detecting efficiency measurement, we use two scintillators to trigger the system as shown in Fig. 5. The detecting efficiency is defined as the proportion of the events that both ends of the strip covered by the scintillators have valid signals

\begin{center}
\includegraphics[width=7cm]{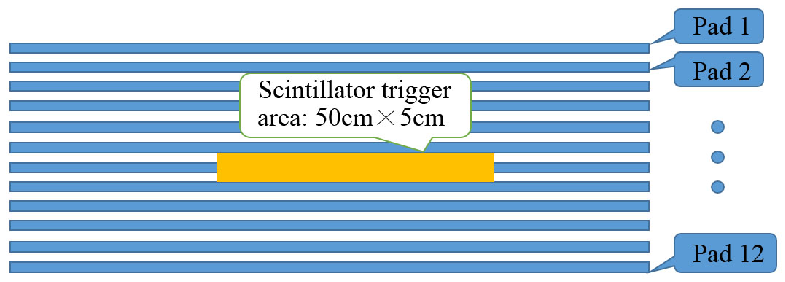}
\figcaption{\label{fig2}Experimental setup for measuring the detecting efficiency.}
\end{center}

\subsubsection{TOT correction}

As signals are read out at each end of the strips, the mean time $T_{ave}=(T_{end1}+T_{end2})/2$ is independent of the hit position along the strip. The absolute flight time of the cosmic ray between top and bottom MRPC is $T_{tof}=T_{ave4}-T_{ave1}$. The events will be selected and analyzed when the muons hit the same strip of all of the four MRPCs. Fig. 6 left panel shows the distribution of $T_{tof}$ on one of the 12 strips without any correction.

\begin{center}
\includegraphics[width=8cm]{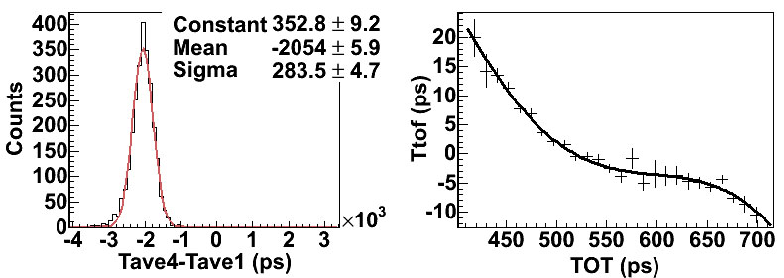}
\figcaption{\label{fig4}Left panel: Time-of-flight without any corrections. Right panel: The time-of-flight between top and bottom module as a function of the pulse width (TOT) of the bottom module. A similar slewing correction can be obtained for the top chamber.}
\end{center}

Two corrections are implemented to calculate the time resolution. The first one is time-slewing correction, as there is a shift in time depending on the pulse height of the input signal, which is encoded into the width of time over threshold (TOT) [7]. A typical TOT distribution with a polynomial-fit curve is shown in Fig. 6 right panel for the bottom MRPC. A similar TOT correction is also made for the top MRPC module.
\\
\subsubsection{Path length correction}

Since the absolute time-of-flight is dependent on the path length (L) of the muon, an additional time-path length correction is needed. The path length depends on the incident angle of muon, and can be calculated via the ¡®hit¡¯ positions along the strips at each module. To determine the ¡®hit¡¯ position along the strip, the velocity of the signal travelling along the strip is necessary.
\\

By using the scintillators to trigger the system at different positions and comparing them with the time differences of the signals from each end $T_{diff}=T_{end1}-T_{end2}$, the velocity of the signal travelling along the strip can be calculated as $v=2\times p0=16.86$ cm/ns [8], where $p0$ is the slope of the time difference $T_{diff}$ versus the triggered position as shown in Fig. 7.

\begin{center}
\includegraphics[width=8cm]{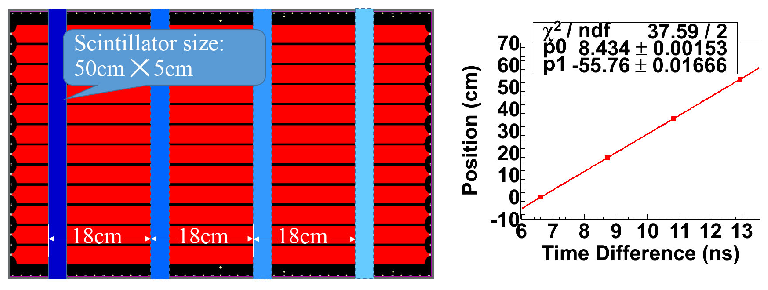}
\figcaption{\label{fig3}Left panel: The position of the scintillator for trigger. Right panel: The time difference of the signals from each end versus triggered position. The slope (p0) gives the velocity of the signal travelling along the strip.}
\end{center}

Then, with the hit position along the strip calculated by the time difference $T_{diff}$, the path length is converted into cm by $L=\sqrt{d^2+[(T_{diff4}-T_{diff1})\times v/2]^2 }$, where $d = 58$ cm is the vertical distance between top and bottom module. The time-path length correlation is shown in Fig. 8 left panel. The line indicates the linear fit to the data.

\begin{center}
\includegraphics[width=8cm]{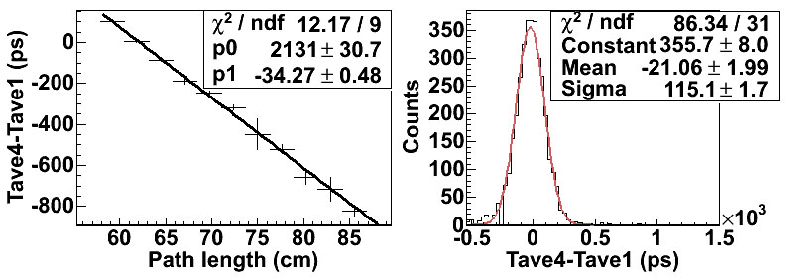}
\figcaption{\label{fig5}Left panel: Variation of time-of-flight measured from the top to the bottom MRPC with the path length. Right panel: Time-of-flight between top and bottom chambers after correction for TOT and path length.}
\end{center}

The slope of this fit can be converted into the average flight velocity of the muons, which is calculated as $1/(34.3$ ps/cm$) = 2.92\times10^8$ m/s. After the corrections, the distribution of the time-of-flight between two MRPC modules ($T_{tof}$) is shown in Fig. 8 right panel. The time resolution for each module is obtained as  $\sigma_t=\sigma_{T_{tof}}/\sqrt{2}=81.3$ ps.

\subsection{The position resolution}

The position resolution along the strip direction is calculated by using the top and bottom MRPCs to predict the position of the `hit' in the middle two MRPCs. The hit position of each module can be calculated by the time difference as $P_i=T_{diff_i}\times v/2$. As the four modules are placed with the same spacing, the predicted hit position of the third module is $P_{MRPC3}=1/3\times P_1+2/3\times P_4$. A typical measurement of the difference between measured and predicted hit position of the third module is shown in Fig. 9. The position resolution of each MRPC should be $\sigma_p=\sigma/\sqrt{1+(2/3)^2+(1/3)^2}=5.3$ mm.

\begin{center}
\includegraphics[width=6cm]{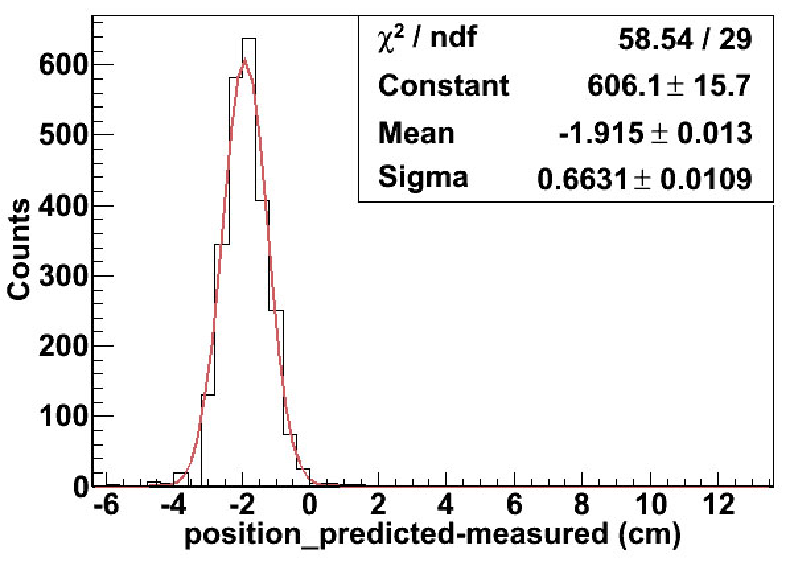}
\figcaption{\label{fig6}Difference of the predicted hit position in the third MRPC with the position obtained from the time difference.}
\end{center}

\section{Conclusions}

In this study we build a large area test platform with four MRPC modules for cosmic ray tracking. The DAQ system of the test platform is triggered when the muon go through all of the four modules. Analysis of the experimental data shows that the test platform has a promising performance for the cosmic ray tracking. The time resolution of the MRPC module is $\backsim$80 ps and the position resolution along the strip can reach 5.3 mm. The measurement of the position resolution is analyzed only along the strip direction. The resolution perpendicular to the strip can be calculated by the distance between the middle of a strip to the next(44 mm), which obey rectangular distribution, thus $\sigma=44/\sqrt{12}$ mm$=12.7$ mm. The time and spatial resolution of the platform can be calculated as $\sigma_{platform}=\sigma_{module}/\sqrt{n} $, where $n$ is the number of modules used. In condition of this, this large area cosmic ray test platform constructed by four MRPC modules can track the cosmic ray with a spatial resolution $\backsim$6.3 mm, and provide a reference time $\backsim$40 ps at the same time.

\section{Acknowledgments}

We would like to acknowledge Dr. H.L.Dai and Dr. X.S.Jang for their support of the readout electronics.

\end{multicols}

\vspace{10mm}

\vspace{-1mm}
\centerline{\rule{80mm}{0.1pt}}
\vspace{2mm}

\begin{multicols}{2}

\end{multicols}

\clearpage


\begin{thebibliography}{90}

\vspace{3mm}

\bibitem{lab1}
M.C.S Williams, E. Cerron et al.,Nucl. Instrum. Methods A, 1999, 434,362-372

\bibitem{lab2}
Y.J.Sun et al., Nucl. Instrum. Methods A, 2008, 593,307-313

\bibitem{lab3}
Sun Yong-Jie et al., Chinese Physics C, 2011, V35N9, 838-843

\bibitem{lab4}
C. Yang, X.J. Huang, C.M. Du et al., arXiv:1402.1078

\bibitem{lab5}
F. Anghinolfi, P. Jarron, A.N. Martemiyanov, et al., Nucl. Instr. Meth., 2004, A533,183

\bibitem{lab6}
F. Anghinolfi, P. Jarron, F. Krummenacher et al., NINO, an ultra-fast, low-power, front-end amplifier discriminator for the Time-Of-Flight experiment in ALICE, presented at 2003 Nuclear Science Symposium, Portland, Oregon.

\bibitem{lab7}
ALICE Collaboration, Addendum to the Technical Design Report of the Time-Of-Flight System, CERN/LHCC/2002-16

\bibitem{lab8}
LI Cheng et al., HEP$\&$NP, 2006, 30(7) (in Chinese)

\end{thebibliography}
\end{document}